\documentstyle[11pt,moriond,epsfig]{article}

\def\rn{}
\def\nn#1 #2{#2. #1}				% Name with 1 initial
\def\nnn#1 #2 #3{#2. #3. #1}			% Name with 2 initials
\def\nnnn#1 #2 #3 #4{#2. #3. #4 #1}		% Name with 3 initials
\def\nnnnn#1 #2 #3 #4 #5{#2, #3. #4. #5. #1}	% Name with 4 initials
\def\dualand{ and\hbox{ }}				
\def\multiand{ and \hbox{ }}				
\def\rf#1;#2;#3;#4;#5 {{\frenchspacing\par\rn#1, {\it #3} {\bf #4}, #5 (#2). \par}}
\def\rfbook#1;#2;#3;#4;#5 {{\frenchspacing\par\rn#1, {\it #3} (#5, #4, #2).\par}}
\def\rfprep#1;#2;#3 {{\par\frenchspacing\rn#1, preprint #3 (#2).\par}}
\def\expec#1{\langle#1\rangle}

\def\etal{{\frenchspacing\it et al.}}
\def\ie{{\frenchspacing\it i.e.}}

\def\beq#1{\begin{equation}\label{#1}}
\def\eeq{\end{equation}}
\def\beqa#1{\begin{eqnarray}\label{#1}}
\def\eeqa{\end{eqnarray}}
\def\eq#1{equation~(\ref{#1})}

\def\spose#1{\hbox to 0pt{#1\hss}}
\def\simlt{\mathrel{\spose{\lower 3pt\hbox{$\mathchar"218$}}
     \raise 2.0pt\hbox{$\mathchar"13C$}}}
\def\simgt{\mathrel{\spose{\lower 3pt\hbox{$\mathchar"218$}}
     \raise 2.0pt\hbox{$\mathchar"13E$}}}
\def\simpropto{\mathrel{\spose{\lower 3pt\hbox{$\mathchar"218$}}
     \raise 2.0pt\hbox{$\propto$}}}

\def\ed{\end{document}}

\font\bfmath=cmmib10
\def\vmu{\hbox{\bfmath\char'026}}	% Bold-face $\mu$

\def\dl{\delta_l}

\def\C{{\bf C}}

\def\zbar{{\bar z}}
\def\dz{\Delta z}
\def\Om{\Omega_m}
\def\Ol{\Omega_\Lambda}
\def\dOm{\Delta\Om}
\def\dOl{\Delta\Ol}
\def\dl{d_L}
\def\dm{\Delta m}
\def\F{{\bf F}}
\def\m{{\bf m}}
\def\tr{\hbox{tr}\>}

\begin{document}

%\vspace*{7.05cm}
\vspace*{7.30cm}
\epsfxsize=3.3truecm\centerline{\epsfbox{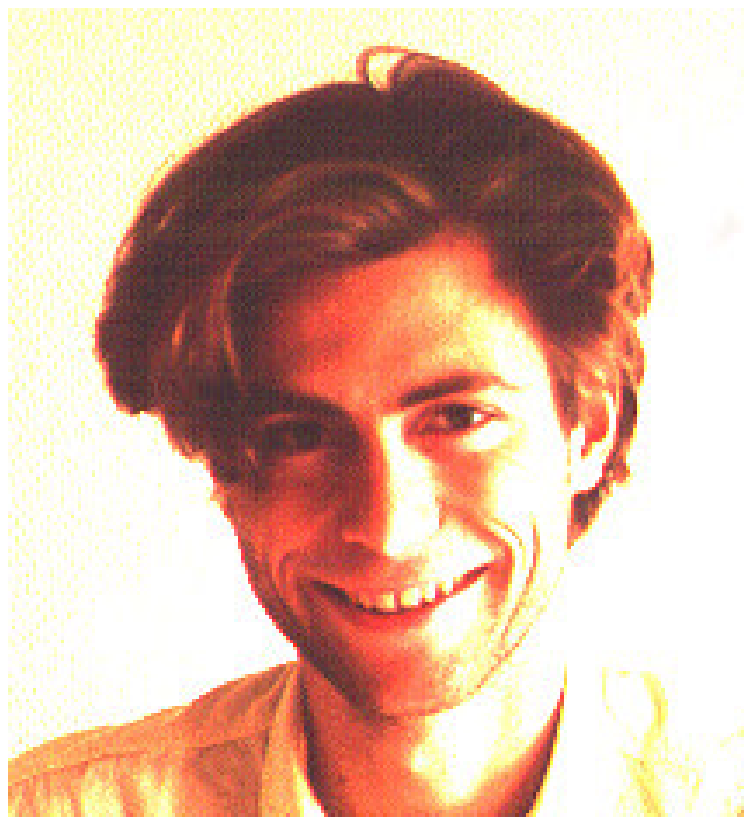}}
%\vskip-11.08cm
\vskip-11.33cm
\vspace*{4cm}
\title{COSMIC COMPLEMENTARITY:\\ COMBINING CMB AND SUPERNOVA
OBSERVATIONS
}

\author{MAX TEGMARK\footnote{Hubble Fellow}, DANIEL J. EISENSTEIN
\& WAYNE HU\footnote{Sloan Fellow}}

\address{Institute for Advanced Study, Princeton, 
NJ 08540\\
max@ias.edu, eisenstein@ias.edu, whu@ias.edu}

\maketitle\abstracts{
We compute the accuracy with which $\Om$ and $\Ol$ can be measured
by combining future SN Ia and CMB experiments, deriving a 
handy expression for the SN Ia Fisher information matrix.
The two data sets are found to be highly complementary: a joint analysis 
reduces the error bars by more than an order of magnitude compared to a
separate analysis of either data set.
}

\section{Introduction}

It may be possible to measure cosmological 
parameters with great accuracy using upcoming
cosmic microwave background (CMB) experiments
\cite{Jungman,parameters,Zalda}, 
galaxy surveys 
\cite{galfisher,Goldberg,neutrinos}
and supernova Ia searches \cite{Goobar,Perlmutter,Garnavich}. 
However, no single type of measurement alone can constrain
all parameters, as it will inevitably suffer from
so-called {\it degeneracies} in which particular combinations of
changes in parameters leave the result essentially unaffected
\cite{parameters,Zalda,confusion,Metcalf}.
Fortunately, different types of cosmological measurements 
are often highly complementary, breaking each other's
degeneracies and combining to give much more accurate measurements
than any one could give alone. 
For example, CMB measurements are highly complementary to 
galaxy surveys \cite{galfisher,neutrinos,Gawiser,Webster,complementarity}.
The topic of this paper is the well-known\cite{Zalda,galfisher,White} 
complementarity between
CMB and SN Ia for measuring the large-scale geometry of spacetime,
given by the density parameters $\Om$ for matter and 
$\Ol$ for vacuum density (cosmological constant).

The accuracy with which
$\Om$ and $\Ol$ can be measured from a SN Ia survey was
first computed by Goobar \& Perlmutter \cite{Goobar}
and subsequently by making $\chi^2$-fits
to real data \cite{Perlmutter,Garnavich,White}.
Here we present the first calculation
of the {\it Fisher information matrix} $\F$ 
for SN surveys, which has the advantage of explicitly
showing how the accuracy depends on the survey details.
We then compare this with CMB information.

\section{The supernova Fisher Matrix}

The Fisher information matrix \cite{karhunen}
quantifies the information content about $\Om$ and $\Ol$.
$\F^{-1}$ gives the best attainable $2\times 2$ covariance 
matrix for the measurement errors on these parameters,
illustrated by the error ellipses in Figure 1b.
$\F$ also allows the SN Ia results to be combined with those from the
CMB, since if independent experiments are analyzed jointly, their
Fisher information matrices simply add.

Suppose that $N$ type Ia supernovae at redshifts $z_1,...,z_N$
have been been observed to have magnitudes
$m_1,...,m_N$. These measurements can be modeled as \cite{Goobar}
\beq{mEq}
m_n = 5\log_{10}[H_0\dl(z_n,\Om,\Ol)] + m_0 + \epsilon_i,
\eeq
where $m_0$ is a constant independent of 
$\Om$, $\Ol$ and $H_0$ and $\epsilon_n$ is a random term with zero
mean ($\expec{\epsilon_n}=0$)
including measurement errors, errors in extinction correction 
and intrinsic scatter in the ``standard candle'' luminosity.
The luminosity distance is 
\beq{dlEq}
H_0\dl = (1+z){S(\kappa I)\over\kappa},\quad
I(z;\Om,\Ol)
=\int_0^z A(z')^{-1/2}dz',
\eeq
\beq{Aeq}
A(z)\equiv (1+z')^2(1+\Om z')-z'(2+z')\Ol,
\eeq
where $S(x)\equiv\sinh x$, $x$ and $\sin x$ for open $(\Om+\Ol<1)$,
flat $(\Om+\Ol=1)$ and closed $(\Om+\Ol>1)$ universes, respectively.
$\kappa\equiv \sqrt{|1-\Om-\Ol|}$.
%%% $\kappa\equiv 1$ for the flat case, and 
%%% $\kappa\equiv \sqrt{|1-\Om-\Ol|}$ otherwise.
Grouping the measured data $m_n$ into an $N$-dimensional vector $\m$
and assuming that the errors $\epsilon_n$ have a Gaussian distribution,
the Fisher matrix is given by \cite{karhunen}
\beq{GaussianFeq}
\F_{ij} = {1\over 2}\tr[\C^{-1}\C,_i\C^{-1}\C,_j] + \vmu,_i^t\C^{-1}\vmu,_j,
\eeq
where $\vmu\equiv\expec{\m}$ is the mean and
$\C\equiv\expec{\m\m^t}-\vmu\vmu^t$ is the covariance matrix of $\m$.
Commas denote derivatives, so 
$\vmu,_i\equiv\partial\vmu/\partial\Omega_i$, $i=m$ or $\Lambda$.
For simplicity, we will assume that 
\beq{Ceq}
\C_{ij}=\delta_{ij}(\dm)^2,
\eeq
\ie, that all the magnitude errors 
$\epsilon_i$ are 
uncorrelated and have the same standard deviation $\dm$,
including systematic errors.
Our treatment below is readily generalized to arbitrary error 
models $\C$. Since $\C,_i=0$, all the information 
about $\Om$ and $\Ol$ comes from the second term in \eq{GaussianFeq}.
Differentiating \eq{mEq} thus gives
\beq{Feq2}
\F_{ij} = {1\over\dm^2}\sum_{n=1}^N w_i(z_n) w_j(z_n), 
%{\partial I\over\partial\Omega_i}(z_n)
%{\partial I\over\partial\Omega_j}(z_n),
\eeq
where 
\beq{wEq}
w_i(z) \equiv \left({5\over\ln 10}\right)
\left\{{\kappa S'[\kappa I(z)]\over S[\kappa I(z)]}
\left[{\partial I\over\partial\theta_i} - {I(z)\over  2\kappa^2}\right] 
+ {1\over  2\kappa^2}
\right\},
\eeq
\beq{PartialEq}
{\partial I\over\partial\Om}(z) 
  = -{1\over 2}\int_0^z{z'(1+z')^2\over A(z')^{3/2}}dz',
\qquad
{\partial I\over\partial\Ol}(z) 
  = {1\over 2}\int_0^z{z'(2+z')\over A(z')^{3/2}}dz'.
\eeq

%\goodbreak

{\footnotesize
\begin{table}[t]
%\caption{
%\parbox{\textwidth}{
Table 1:
Attainable error bars $\Delta\Omega_i$ for various combinations of data
sets.  The rows correspond to using CMB alone and three forecasts
(pessimistic, middle-of-the-road, and optimistic) for available SN Ia
data in five years time.
% The magnitude errors $\dm$ include systematics.
The CMB columns correspond to the upcoming MAP and Planck satellite
missions without ($-$) and with $(+)$ polarization information.
Planck$+$ is seen to improve over the ``No CMB'' column by over an
order of magnitude, and the difference is even greater between the
``Opt'' and ``No SN'' rows.  
The ``No SN'' row is overly conservative, since gravitational lensing 
breaks the CMB degeneracy somewhat$^{11}$
%\cite{Metcalf}, 
but this lensing information is 
dwarfed by the SN Ia in the other rows. 
%}
%\vspace{0.4cm}
\vspace{0.3cm}
\begin{center}
\begin{tabular}{|l|cccc|cc|cc|cc|cc|cc|}
\hline
&&&&&\multicolumn{2}{c|}{No CMB}&\multicolumn{2}{c|}{MAP$-$}&\multicolumn{2}{c|}{MAP$+$}&\multicolumn{2}{c|}{Planck$-$}&\multicolumn{2}{c|}{Planck$+$}\\
SN Ia	&$N$	&$\dm$	&$\zbar$&$\dz$	&$\dOm$	&$\dOl$	&$\dOm$	&$\dOl$	&$\dOm$	&$\dOl$	&$\dOm$	&$\dOl$	&$\dOm$	&$\dOl$	\\
\hline	
No SN	&0	&-	&-	&-	&$\infty$&$\infty$&3.6	&3.2	&2.0	&1.8	&3.0	&2.6	&.63	&.54	\\
Pess	&100	&0.5	&0.55	&0.2	&.81	&1.1	&.12	&.12	&.12	&.10	&.12	&.10	&.11	&.10	\\
Mid	&200	&0.3	&0.65	&0.3	&.22	&.34	&.06	&.08	&.05	&.05	&.05	&.04	&.04	&.04	\\
Opt	&400	&0.2	&0.70	&0.4	&.08	&.14	&.04	&.07	&.02	&.03	&.02	&.02	&.02	&.02	\\
\hline
\end{tabular}
\vspace{-0.2cm}
\end{center}
\end{table}
}

\begin{figure}[h]
%\figone
\epsfxsize=16.0truecm\epsfbox{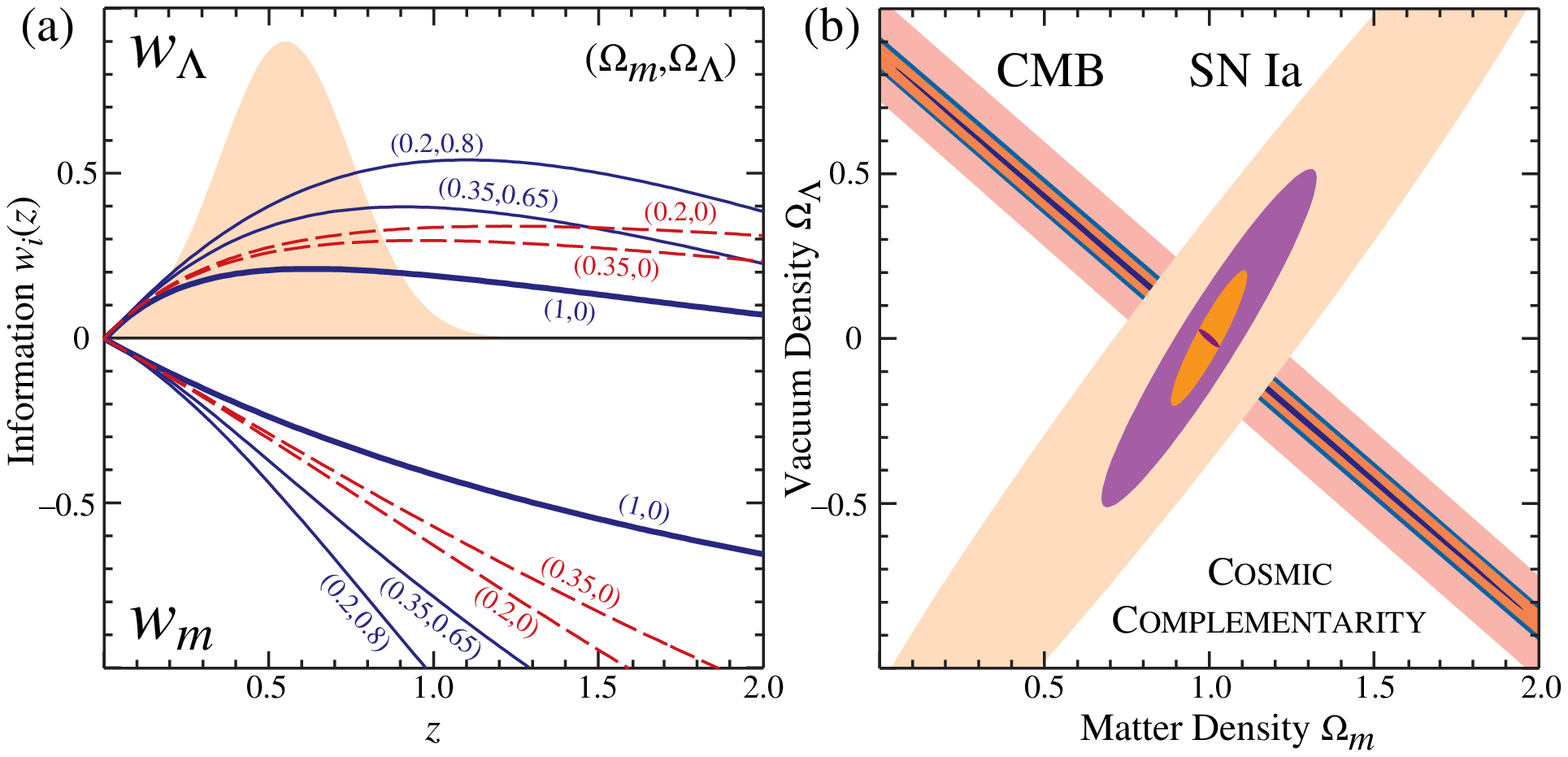}
%\caption{
{\footnotesize
Figure 1: Figure 1a (left) shows the weight functions $w_\Lambda$ (positive)
and $w_m$ (negative) for standard CDM,
two open $(\Ol=0)$ models and two flat $(\Ol=1-\Om)$ models. 
The Fisher matrix element $\F_{ij}$ is  
computed by simply integrating the product of the curves $w_i$ and $w_j$ and a
redshift distribution $f$ such as the shaded one.
Figure~1b (right) shows the $68\%$ confidence regions obtained 
from the three hypothetical SN Ia data
sets specified in Table 1, which are intended to 
be best-case, middle-of-the-road and worst-case
scenarios for what might be available in five years time.
Four corresponding ellipses for upcoming CMB 
experiments are also shown, based on a full 12-parameter 
analysis described elsewhere$^{14}$. The assumed fiducial model is 
COBE-normalized CDM with $\Om=1$, $\Ol=0$, $\Omega_b=0.05$, 
$\Omega_\nu=0.05$ and $h=0.5$.
%\protect{\cite{complementarity}}. 
Combining  the CMB and SN Ia data shrinks the error region to the 
overlap of the two corresponding ellipses: 
for instance, the tiny black ellipse in the center if for a 
joint analysis of the optimistic SN Ia case with polarized Planck data.
%\label{FigOne}
}
\end{figure}

\noindent
The expression in braces approaches $I^{-1} \partial I/\partial\theta_i -I^2/6$ as $\kappa\to 0$.
It is instructive to rewrite
\eq{Feq2} as
\beq{Feq3}
\F_{ij} = {N\over(\dm)^2}\int_0^\infty f(z) w_i(z) w_j(z) dz,
\eeq
where
the SN Ia redshift distribution is given by
$f(z) = {1\over N}\sum_{n=1}^N \delta(z-z_n)$.
The contribution to $\F$ from each redshift can thus be split
into two factors, one  
reflecting the quality of the data set ($N f[z]/\dm^2$)
and the other incorporating the effects of cosmology
(the weight functions $w_i$). 
The functions $w_i$ are plotted in Figure 1a for a variety of
cosmological models.  If all the observed supernovae were at the same
redshift $z$, then the resulting $2\times 2$ Fisher matrix
$\F_{ij}\propto w_i(z) w_j(z)$ would have rank 1, \ie, be singular.
The vanishing eigenvalue would correspond to the eigenvector
$(w_\Omega,-w_\Lambda)$.  Physically, this is because there is more
than one way of fitting a single measured quantity $d_L(z)$ by varying
two parameters ($\Om$ and $\Ol$).  The corresponding ellipse in Figure
1b would be infinitely long, with slope $-w_\Omega/w_\Lambda$, the
ratio of the magnitudes of the $\Om$ and $\Ol$ curves in Figure 1a at
that redshift.  The SN Ia ellipses plotted in Figure 1b correspond to a
range of redshifts, with $f$ being a Gaussian of mean $\zbar$ and
standard deviation $\dz$ given by Table~1. This breaks the degeneracy
only marginally, leaving the SN ellipses quite skinny, since the ratios
$w_\Omega/w_\Lambda$ in Figure 1a are seen to vary only weakly with
$z$.

\section{Conclusions}

In conclusion, we have derived a  handy expression for 
the SN Ia Fisher information matrix and combined it with the Fisher matrix of the CMB.
Whereas two identical data sets only give a measly factor of $\sqrt{2}$ improvement in 
error bars when combined, the gain factor was found to exceed 10  in this case.
This ``cosmic complementarity'' is due to the fortuitous fact that 
although either data set alone suffers from a 
serious degeneracy problem, the directions in which they 
are insensitive (in which the
ellipses in Figure 1b are elongated) are almost orthogonal.
This is because the CMB probes the redshift-distance relationship via the location of the Doppler peaks,  
depending mainly on $\Om+\Ol$ for standard CDM, whereas the SN Ia probe the redshift-luminosity
relation, measuring roughly $\Om-\Ol$. The complementarity remains just as striking for the 
more general cosmological models plotted in Figure 1a---both the CMB 
and SN Ia ellipses simply rotate somewhat in these cases,
but remain very skinny and almost perpendicular.

The potential power of upcoming CMB measurements has led to a widespread feeling that they
will completely dominate cosmological parameter estimation, leaving other types
of experiments making only marginal contributions.
Because of cosmic complementarity, of which the present paper gives but one example
out of many, this view is misleading:
two data sets combined can be much more useful than either one alone.

\section*{Acknowledgments}
We thank Alex Kim, Saul Perlmutter, Adam Riess and Martin White for 
useful discussions. 
MT was supported by NASA though grant NAG5-6034 and a Hubble Fellowship,
HF-01084.01-96A, awarded by STScI, which is operated by AURA, Inc. 
under NASA contract NAS5-26555. 
DJE and WH were supported by NSF PHY-9513835. WH was additionally
supported by a Sloan Fellowship and the W. M. Keck Foundation.

%%%%%%%%%%%%%%%%%%%%%% REFERENCES: %%%%%%%%%%%%%%%%%%%%%%%%%

\end{document}